# Curved Gratings as Plasmonic Lenses for Linearly Polarised Light

*Alireza Maleki[1,2\*], Thanh Phong Vo[1,2], Antoine Hautin[1,2,3],*

*James E Downes[2], David W Coutts[1,2], and Judith M Dawes[1,2]*

[1]ARC Centre of Excellence CUDOS, Macquarie University, Sydney NSW 2109, Australia
[2]MQ Photonics Research Centre, Dept. of Physics and Astronomy, Macquarie University, Sydney NSW 2109, Australia
[3]Ecole Centrale de Lyon, 36 Avenue Guy de Collongue, 69134 Ecully Cedex, France
\*alireza.maleki@mq.edu.au

Abstract: The ability of curved gratings as sectors of concentric circular gratings to couple linearly polarized light into focused surface plasmons is investigated by theory, simulation and experiment. Curved gratings, as sectors of concentric circular gratings with four different sector angles, are etched into a 30-nm thick gold layer on a glass coverslip and used to couple linearly-polarised free space light at 700 nm into surface plasmons. The experimental and simulation results show that increasing the sector angle of the curved gratings decreases the lateral spotsize of the excited surface plasmons, resulting in focussing of surface plasmons which is analogous to the behaviour of classical optical lenses. We also show that two faced curved gratings, with their groove radius mismatched by half of the plasmon wavelength (asymmetric configuration), can couple linearly-polarised light into a single focal spot of concentrated surface plasmons with smaller depth of focus and higher intensity in comparison to single-sided curved gratings. The major advantage of these structures is the coupling of linearly-polarised light into focused surface plasmons with access to and control of the plasmon focal spot, which facilitates potential applications in sensing, detection and nonlinear plasmonics.

## 1. Introduction

Engineered metallic nano-structures have the ability to confine and manipulate electromagnetic waves as collective oscillations of free electrons at a metal-dielectric interface at nanometre length scales, known as surface plasmons [1]. The capability of surface plasmons to concentrate electromagnetic fields with high local field intensities has attracted attention for novel nano-photonic technologies in sensing [2-4], photo-detection [5-7], nonlinear optics [8-10], and nano-photonic devices and circuitry [11-13]. Surface plasmons also show a larger effective propagation constant than light, necessitating specific techniques for coupling photons to surface plasmons. The coupling methods are in principal based on either photon tunnelling, for example, total internal reflection by prisms, or diffraction via tailored plasmonic structures, such as gratings and nano-antennas.

Plasmonic gratings have emerged as one of the best practical techniques to efficiently couple photons into surface plasmons on 2D platforms [14]. In addition, because of the highly confined electric field of plasmons to the metal-dielectric interface, tailoring the geometry of the gratings provides the opportunity to manipulate the surface plasmons. Here we show the capability of curved gratings, as sectors of full circular gratings, to focus the coupled incident linearly polarised (*p*-polarised) light into surface plasmons.

Plasmonic focusing concentrates surface plasmons in 2D at a metal-dielectric interface. Various plasmonic focusing structures have been demonstrated experimentally, including engineered arrays of metallic nano-slits [15], an arc of nano-holes [16], parabolic nanoparticle chains [17], diffraction gratings [18], and an elliptical corral [19] with particular applications in wavelength division multiplexing, spectral filtering, waveguide coupling, correction of divergent surface plasmon beams and angular interferometry, respectively. Elliptically curved gratings [20] and periodically corrugated wires have been proposed and investigated for terahertz frequencies [21].

Single circular slits can also act as surface plasmon focusing elements [22-24]. However, using linearly-polarised light results in two high intensity spots along the diameter of the circular slits in the polarization direction of the beam, due to destructive interference of counter-propagating surface plasmons at the centre of the circle. In order to have a single spot at the centre of a circular slit one needs to illuminate the structure with radially-polarised light for constructive interference of counter-propagating surface plasmons at the centre of the circular slit. Further, it has been shown that circular gratings, consisting of several circular slits, can focus surface plasmons to the centre of the concentric circles when illuminated with circularly-polarised light [25-27]. Due to the increased number of slits in circular gratings the focused surface plasmons have a higher intensity in comparison to a single circular slit. Linearly-polarised light may also be coupled to focus surface plasmons by placing two concentric half-circular slits or corrals facing each other. However, the slit radii need to be mismatched by half of the plasmon wavelength, to provide phase matching and enable constructive interference of the counter-propagating surface plasmons at the centre of the gratings [28,29].

Curved gratings are sectors of full concentric circular gratings, as illustrated in Fig. 1. Here we investigate the ability of curved gratings to focus surface plasmons. A key requirement of plasmonic focusing elements is the ability to control the width of the lateral distribution of the coupled surface plasmons. We present simulations and experimental results that demonstrate the effect of the sector angle of curved gratings on the width of the focussed surface plasmons, when they are illuminated with linearly-polarised light (*p*-polarised). In comparison, full circular gratings offer efficient focusing of surface plasmons to the grating centre when radially-polarised light illumination is well-aligned with the radial centre of the circular slits, which can be challenging in practice. Furthermore, curved gratings are positioned to one side of the focal spot and the region beyond the focal point allows access to concentrated surface plasmons for further processes. An alternative design uses two facing curved gratings with identical sector angles in an asymmetric concentric configuration (with the corresponding groove radii mis-matched by half of the surface plasmon wavelength). This design can focus surface plasmons to a smaller longitudinal spotsize, with higher intensity, in comparison to the single sector curved gratings. Considering these advantages, these asymmetric double curved grating structures have potential applications in sensing, detection and nonlinear plasmonics.

**Fig. 1** SEM micrograph of fabricated curved gratings for four different sector angles

## 2. Fabrication and experimental results

Four different sector angles $20°$, $40°$, $120°$, and $180°$, each with seven grooves, were selected for the fabrication of the curved gratings. The inner groove radius was selected so as to maintain a constant effective aperture size of $3.0\ \mu m$ for each structure. The 30-nm-thick gold films were deposited on a glass substrate (refractive index $n = 1.5$) by ion-assisted deposition using an oxygen plasma-assisted thermal deposition system, to ensure good adhesion, flatness and uniformity of the gold film on the glass [30]. The curved gratings were fabricated with grooves milled through the full thickness of the gold layer. Focused ion beam (FIB) etching, using a Zeiss Auriga 60 cross beam scanning electron microscope (SEM) with an Orsay Canion FIB column, was used to mill the grooves into the gold layer. Fig. 1 shows SEM micrographs of the fabricated curved gratings. Characterization of the fabricated structures by SEM shows the final grating period $\Lambda_{gr} = 770 \pm 10$ nm with groove width (*GW*) of half of the grating period ($GW = \dfrac{\Lambda_{gr}}{2} = 385$ nm).

The performance of the fabricated curved gratings is studied by near-field scanning optical microscopy (NSOM) with an appropriate optical coupling system as shown schematically in Fig. 2. The excitation light had a wavelength of 700 nm with ∼5 nm bandwidth, illuminating the grating by a $10\times$ microscope objective from below the sample through the glass substrate. The NSOM tip was positioned at a height of 10 nm from the gold surface to collect the evanescent field of the surface plasmons excited by the gratings.

**Fig. 2** (a) Schematic of NSOM experimental setup

**Fig. 3** NSOM scan (a) compared with simulation intensity map, (b) showing the focusing of coupled surface plasmons for $120°$ sector angle curved gratings

In fig. 3(a), the NSOM image ($256 \times 256$ pixels) for a $120°$ sector angle grating shows the focused surface plasmons. There is a phase change in the image, which is attributed to irregularities of the handmade NSOM probe. 3D simulations of the curved gratings were also performed using finite element methods (FEM) and adaptive meshing by COMSOL. The simulation parameters were identical to the fabricated structure and experiment parameters. Fig. 3 (b) shows the near-field intensity map at 10 nm vertical height above the gold-air interface, which demonstrates the focusing of surface plasmons in good agreement with NSOM scans. Focusing is also evident in the near field intensity profiles along the bisector of the gratings (along the horizontal dashed cutline in fig.3 (b)) for both the NSOM scan (black solid line) and the 3D simulation (red dashed line) as shown in Fig. 4. The intensity profiles are normalized to the maximum value of intensity of each line. In both cases, scattering of the light at the grooves contributes to observed oscillations in the positive x-coordinate and the focal spot occurs at the zero position. The results show the concentration of surface plasmons at the focal spot, although the field is relatively weak. This weak field enhancement is due to the off resonance of the illumination wavelength together with the 30 nm depth of grooves. A very thin gold layer results in coupling of some surface plasmons onto the gold-glass interface. Later simulation optimization showed that the most efficient coupling and consequently the best field enhancement of the focal spot would occur for 120 nm gold thickness (groove depth) when illuminated at the peak resonance wavelength ($\lambda_0 = 790$ nm) for the grating period of $\Lambda_{gr} = 770$ nm.

**Fig. 4** Intensity profile along the bisector (x axis in fig.6) of the simulated and experimental gratings (both with grating period of $\Lambda_g = 770$ nm) determined at a 10 nm vertical height from the gold-air interface

The width of the focal spot is defined by the full width at half maximum (FWHM) of the intensity profile along the normal to the bisector of the gratings passing through the peak of the focal spot (along the vertical dashed cutline in Fig. 3(b)). Fig. 5 compares the simulation calculations and NSOM measurement of the width of the focal spot for different sector angles. With increasing sector angle, the width of the coupled surface plasmons is decreased; for sector angles over $100°$ the coupled surface plasmons are well focused at the focal spot with a minimum spot size of 300 nm, which is $(\frac{\lambda_\circ}{2.3})$. The decreasing width of the focal spot with increasing sector angle is analogous to the performance of classical optical lenses. These plasmonic lenses can couple and direct surface plasmons into the focal spot allowing access to the focal spot for further processing.

**Fig. 5** Measurement and simulation for the vertical width of the focal spot from simulation (blue diamonds) and experimental measurement (red circles), and reciprocal of the effective NA (black line) which is consistent with the experimental and simulation results

In contrast to the excitation of linear gratings, for curved gratings, the *p*-polarized light makes different angles with the normal to the grooves at each point, as shown schematically in fig. 6. As a result, the amplitude of the surface plasmons arising from each point of the grooves drops as the cosine of the angle between the light polarization and the grating vector ($\cos(\alpha)$), [31]. Consequently the total intensity of the coupled surface plasmons through the aperture of the grating is proportional to;

$$I_{sp} \propto \int_{-\beta}^{\beta} \cos(\alpha)^2 \, d\alpha \qquad (1)$$

Where $\beta$ is half the sector angle of the curved grating, and $\alpha$ is the angle between the plane of the linearly polarised beam (*p*-polarized) and the normal to the grooves at each point, as shown in Fig. 6.

**Fig. 6** Schematic of curved grating with sector angle of $120°$. The linearly polarised light makes an angle of $\alpha$ at each point with the grating grooves

Equation (1) shows that increasing the grating sector angle, increases the coupling of the incident optical field to the surface plasmons by the curved gratings. In addition, the experimental and simulation results show that by increasing the sector angle the width of the focus of surface plasmons decreases. As a result, we define the effective numerical aperture (NA) of the curved gratings as

$$NA = \int_0^\beta \cos(\alpha)^2 \, d\alpha = \frac{\sin(2\beta)}{4} + \frac{\beta}{2} \qquad (2)$$

In order to show the appropriateness of the effective numerical aperture, its reciprocal versus sector angle is plotted in Fig. 5 (solid black line). The inverse matches closely to the simulation and experiment data points of the focal width, analogous to the inverse of the numerical aperture of classical optical lenses.

## 3. Investigating plasmonic grating parameters by simulation

Grooves of a plasmonic grating scatter linearly polarised light, (*p*-polarisation), to provide the required extra momentum for the photons to be coupled into surface plasmons. The efficiency of the scattering is mostly dependent on the groove characteristics – the period, width and depth of the grooves ($\Lambda_{gr}$, *GW* & *GD*). Detailed discussion of plasmonic gratings can be found elsewhere [32,33], so here, we discuss the influence of the width and depth of the grating grooves to optimize curved gratings parameters. While the width of the grooves predominantly affects the phase delay between excited surface plasmons from different grooves, the groove depth influences the scattering [25,32]. To save computer memory and time, we simulate linear gratings, with groove depth equal to the full thickness of the metal layer as shown in Fig.7 (a), to optimise the grating parameters.

**Fig. 7** (a) Trench grating illuminated through the gold substrate (material with higher refractive index), (b) average intensity of coupled incident light into surface plasmons versus the depth of grooves (*DG*). Each colour symbol shows a different groove width (*GW*) as a ration of the grating period ($\Lambda_{gr}/m$)

In order to optimize the effect of groove width and depth for efficient coupling of photons into surface plasmons, 2D simulations of linear trench gratings were calculated using the finite element method (FEM) and adaptive meshing by COMSOL. The selected grating period for the simulation was $\Lambda_{gr} = 770$ nm and the grating was illuminated at normal incidence from below through the substrate (glass with refractive index $n = 1.5$). The illumination wavelength was calculated according to the resonance wavelength of the plasmonic gratings under normal incidence (first order) [32];

$$k_{sp} = \frac{2\pi}{\Lambda_{gr}} \qquad (3)$$

in which $k_{sp}$ is related to the illumination wavelength;

$$k_{sp} = \frac{2\pi}{\lambda_\circ} \left( \frac{\varepsilon_m \varepsilon_d}{\varepsilon_m + \varepsilon_d} \right) \qquad (4)$$

With $\varepsilon_m$ and $\varepsilon_d$ as the dielectric constants of metal (gold) and dielectric (air), respectively.

As a result the grating was illuminated at $\lambda_\circ = 790$ nm with the gold layer dielectric constants ($\varepsilon$) of $\varepsilon = -23.3 + i1.46$ taken from Johnson and Christy [34]. The model geometry was meshed by a maximum element size smaller than $\lambda/8$ for the dielectric and smaller than $\lambda/50$ for the thin metal layer to ensure convergence. Fig. 7 (b) shows the average near-field intensity of surface plasmons calculated along a short cutline. Each coloured line and symbol relates to data for a specific groove width corresponding with a fraction of the grating period.

The selected groove widths for the fabricated structures were equal to half of the period of the gratings. The graph in Fig.7 (b) shows that the most efficient couplings for all the groove depths occur for a groove width of half of the grating period, or in other words, the surface plasmon wavelength ($GW = \frac{\Lambda_{gr}}{2} = \frac{\lambda_{sp}}{2} = 385$ nm). In fact, at this width, the phase change for the excited surface plasmons propagating across the other grooves would be zero, resulting in constructive interference of surface plasmons [25].

In addition, Fig.7 (b) shows that maximum plasmonic coupling takes place at groove depths of 120 nm for all groove widths. The depth of the grooves in the case of trench gratings mostly influences the scattering of the incident light beam. For small depths the scattering is not efficient but for deeper grooves scattering increases. However, very deep grooves would also scatter out the excited surface plasmons as radiation, reducing the coupling. Although the 30 nm groove depth of the fabricated trench curved gratings decreases the efficiency of coupling the incident light into surface plasmons, it does not affect the performance of curved gratings as plasmonic lenses.

The number of grooves also influences the intensity of the coupled surface plasmons. Assuming the phase matching condition is met, increasing the number of grooves of circular gratings increases the coupled surface plasmon intensity. However, by increasing the radius of grooves beyond the propagation length of the surface plasmons the influence of added grooves to the grating decreases and approaches to zero. A detailed discussion of the role of the number of grooves can be found elsewhere [25,27]. Our simulation also shows that for the curved gratings with period equal to the surface plasmon wavelength and groove width of half of the grating period (consistent with phase matching when illuminated at normal incidence), increasing the number of grooves to seven increases the intensity of the focal spot.

Using 3D COMSOL simulations, we also tested the wavelength response of the curved gratings, in order to compare with the resonance wavelength of linear gratings through Eq. (3). The simulated curved grating period is identical to the period of the linear grating and fabricated structures ($\Lambda_{gr} = 770$ nm). The groove width is equal to half of the period of the grating and the thickness of the gold layer (groove depth) is also identical to the thickness of the fabricated curved grating (30 nm). The wavelength response of the grating is illustrated in Fig. 8 with maximum response of the grating at $\lambda_{max} = 780$ nm, very close to the calculated value of resonance wavelength for linear gratings with identical grating period ($\lambda_\circ = 790$ nm), with an estimated bandwidth of 100 nm as the full width at half maximum (FWHM) of the fitted Gaussian curve.

**Fig. 8** Wavelength response of the curved grating from simulations – 30 nm thick gold film on glass with 120 degree sector angle

This simulation shows that the selected wavelength for the illumination of the structure in the experiment is not the resonance wavelength which results in less efficient coupling of incident light into surface plasmons. However, it does not affect the performance of the curved gratings as plasmonic lenses and only results in weaker NSOM signals through near field scanning. The ideal situation of a $120°$ sector angle curved grating with optimized values of structural parameters for grating period of $\Lambda_{gr} = 770$ nm ($GW = \frac{\Lambda_{gr}}{2} = 385$ nm, gold thickness=$GD$=120 nm), and illuminated at $\lambda_\circ = 790$ nm was also simulated, see Fig. 9 (a,b). Fig. 9 (a) shows the intensity map of the near field and Fig. 9 (b) exhibits the intensity profile along the bisector of the grating (along the horizontal dashed cutline). They show the focusing of surface plasmons with higher field enhancement at the focal spot in comparison to Fig 3(b) and Fig. 4, due to the optimized parameters.

**Fig. 9** (a) 3D simulation of near field intensity map of $120°$ sector angle curved grating with optimized structural parameters ($GW$= 385 nm, and $GD$=120 nm) illuminated at $\lambda_0 = 790$ nm, (b) The intensity profile along the bisector of the grating (x axis in Fig. 6)

The focal spot of the curved gratings are not perfectly circular and are stretched to some extent introducing the focus depth. For the polarization direction of the incident light along the bisector of the curved grating ($x$ direction) the depth of focus will be along the $x$ axis. While we tried to perfectly align the direction of polarization of the incident light along the bisector of the curved gratings, simulation shows that by rotating the polarization angle in relative to the curved grating bisector, depth of the focus rotates. Fig. 10 (a) shows the rotation of the depth of focus versus the angle between the polarization direction of the incident light and the bisector of the curved grating up to $50°$. Over $50°$ the focal spot becomes distorted. As it is obvious from the graph, for misalignment angle over $30°$ the rotation of the depth of focus becomes appreciable. Fig. 10 (b) also shows the rotated depth of focus (yellow arrow) when the angle between the polarization direction of the incident light and the curved grating bisector is $40°$ (red arrow).

### 4. Asymmetric faced curved gratings for a tighter focusing of surface plasmons

Although curved gratings focus surface plasmons, the depth of their focal spot is long in comparison to the width of the focal spot. This can be seen in Fig.3 and Fig.4. To address this problem, we studied arrangement of two curved gratings centred at the same point facing each other to focus surface plasmons using linearly polarised light, as shown in Fig.10.

**Fig. 10** Schematic for two faced curved gratings centred at the same point and in a symmetry-broken configuration

**Fig. 11** Simulation of the performance of single curved grating and asymmetric double sided curved gratings with six grooves

The radius of the grooves of each curved grating is mis-matched half of the period of the surface plasmons relative to the opposite grating in order to provide the phase matching condition for counter-propagating surface plasmons to interfere constructively. Fig.11 (a and b) show the 3D simulation of a single curved grating and an asymmetric double-sided curved grating for $160°$ sector angle with $\Lambda_{gr} = 770$ nm, illuminated at $\lambda_\circ = 790$ nm and with groove width of half of the grating period. The oval shape focal spot in single curved grating (fig.11 (a)) changes to a small dot with higher intensity, demonstrating a significant decrease in the longitudinal distribution of plasmons at the focal spot (Fig. 11(b)). A detailed comparison is shown in Fig. 12 in which the intensity profile along the bisector of the gratings is plotted. It shows that the longitudinal width is changed from $800$ nm for the single curved grating to $260$ nm ($\approx \frac{\lambda}{3}$) for the asymmetric double-sided curved grating with a significant enhancement in the intensity of the hot spot in comparison to the focused surface plasmons from the single curved gratings. However, the width of the focal spot does not change.

**Fig. 10** Comparison of the longitudinal intensity profile along the bisector of the single curved grating and asymmetric double-sided curved gratings with six grooves

### 5. Conclusion

The capability of curved gratings to couple linearly polarised light to focused surface plasmons is investigated by theory and simulation and demonstrated experimentally. It is shown that controlling the sector angle of the curved gratings offers an additional tool to manipulate the width of the lateral distribution of surface plasmons. Moreover, curved gratings are positioned on one side of the focused surface plasmon spot, allowing the access to the concentrated surface plasmons for additional processes. Although illuminating a circular grating with radially-polarised light provides a sharp high intensity focal spot, it requires careful alignment of the centres of the beam and the structure, which is challenging in practice. In addition, the closed geometry of the circular gratings would limit their application.

The focal spot of curved gratings is not symmetric: consisting of longer longitudinal focus depth than the width of the focal spot. However, it is shown that facing two curved gratings in an asymmetric configuration and illuminating with linearly polarised light along the bisector of the curved gratings results in a very sharp focal point, similar to the full circular gratings when are illuminated with circularly polarised light. Curved gratings are amenable to planar architectures with potential applications in nonlinear plasmonics, plasmonic detectors and sensing.

## Acknowledgment

This research was supported by the Australian Research Council Centre of Excellence for Ultra-high bandwidth Devices for Optical Systems (project number CE110001018) and Macquarie University. We also acknowledge CSIRO for the FIB facilities and thank Steven Moody for FIB operation.

## References and Links

1. Maier SA (2007) Plasmonics: Fundamentals and Applications: Fundamentals and Applications. Springer,
2. Li XF, Yu SF (2010) Extremely High Sensitive Plasmonic Refractive Index Sensors Based on Metallic Grating. Plasmonics 5 (4):389-394. doi:10.1007/s11468-010-9155-6
3. Dhawan A, Canva M, Vo-Dinh T (2011) Narrow groove plasmonic nano-gratings for surface plasmon resonance sensing. Opt Express 19 (2):787-813. doi:Doi 10.1364/Oe.19.000787
4. Wong C, Olivo M (2014) Surface Plasmon Resonance Imaging Sensors: A Review. Plasmonics 9 (4):809-824. doi:10.1007/s11468-013-9662-3
5. Ishi T, Fujikata J, Makita K, Baba T, Ohashi K (2005) Si nano-photodiode with a surface plasmon antenna. Jpn J Appl Phys, Part 2 44 (12-15):L364-L366. doi:Doi 10.1143/Jjap.44.L364
6. Ren FF, Ang KW, Ye JD, Yu MB, Lo GQ, Kwong DL (2011) Split Bull's Eye Shaped Aluminum Antenna for Plasmon-Enhanced Nanometer Scale Germanium Photodetector. Nano Lett 11 (3):1289-1293. doi:Doi 10.1021/Nl104338z
7. Tang L, Kocabas SE, Latif S, Okyay AK, Ly-Gagnon DS, Saraswat KC, Miller DAB (2008) Nanometre-scale germanium photodetector enhanced by a near-infrared dipole antenna. Nature Photon 2 (4):226-229. doi:DOI 10.1038/nphoton.2008.30
8. Xu T, Jiao X, Zhang G-P, Blair S (2007) Second-harmonic emission from sub-wavelength apertures: Effects of aperture symmetry and lattice arrangement. Opt Express 15 (21):13894-13906
9. Nahata A, Linke RA, Ishi T, Ohashi K (2003) Enhanced nonlinear optical conversion from a periodically nanostructured metal film. Opt Lett 28 (6):423-425
10. Kauranen M, Zayats AV (2012) Nonlinear plasmonics. Nature Photon 6 (11):737-748
11. Bozhevolnyi SI, Volkov VS, Devaux E, Laluet JY, Ebbesen TW (2006) Channel plasmon subwavelength waveguide components including interferometers and ring resonators. Nature 440 (7083):508-511. doi:Doi 10.1038/Nature04594
12. Chen Z-x, Wu Z-j, Ming Y, Zhang X-j, Lu Y-q (2014) Hybrid plasmonic waveguide in a metal V-groove. Aip Adv 4 (1):-. doi:doi:http://dx.doi.org/10.1063/1.4861582
13. Kumar G, Li SS, Jadidi MM, Murphy TE (2013) Terahertz surface plasmon waveguide based on a one-dimensional array of silicon pillars. New J Phys 15. doi:Artn 085031

Doi 10.1088/1367-2630/15/8/085031

14. Lu J, Petre C, Yablonovitch E, Conway J (2007) Numerical optimization of a grating coupler for the efficient excitation of surface plasmons at an Ag-SiO2 interface. J Opt Soc Am B 24 (9):2268-2272. doi:Doi 10.1364/Josab.24.002268
15. Tanemura T, Balram KC, Ly-Gagnon DS, Wahl P, White JS, Brongersma ML, Miller DA (2011) Multiple-wavelength focusing of surface plasmons with a nonperiodic nanoslit coupler. Nano Lett 11 (7):2693-2698. doi:10.1021/nl200938h
16. Gaborit G, Armand D, Coutaz J-L, Nazarov M, Shkurinov A (2009) Excitation and focusing of terahertz surface plasmons using a grating coupler with elliptically curved grooves. Appl Phys Lett 94 (23):231108. doi:10.1063/1.3153125
17. Radko IP, Bozhevolnyi SI, Evlyukhin AB, Boltasseva A (2007) Surface plasmon polariton beam focusing with parabolic nanoparticle chains. Opt Express 15 (11):6576-6582. doi:Doi 10.1364/Oe.15.006576
18. Zhao CL, Wang JY, Wu XF, Zhang JS (2009) Focusing surface plasmons to multiple focal spots with a launching diffraction grating. Appl Phys Lett 94 (11). doi:Artn 111105

Doi 10.1063/1.3100195

19. Drezet A, Stepanov AL, Ditlbacher H, Hohenau A, Steinberger B, Aussenegg FR, Leitner A, Krenn JR (2005) Surface plasmon propagation in an elliptical corral. Appl Phys Lett 86 (7). doi:Artn 074104

Doi 10.1063/1.1870107

20. Gaborit G, Armand D, Coutaz JL, Nazarov M, Shkurinov A (2009) Excitation and focusing of terahertz surface plasmons using a grating coupler with elliptically curved grooves. Appl Phys Lett 94 (23). doi:Artn 231108

Doi 10.1063/1.3153125

21. Maier SA, Andrews SR, Martín-Moreno L, García-Vidal FJ (2006) Terahertz Surface Plasmon-Polariton Propagation and Focusing on Periodically Corrugated Metal Wires. Phys Rev Lett 97 (17):176805
22. Liu ZW, Steele JM, Srituravanich W, Pikus Y, Sun C, X Z (2005) Focusing surface plasmons with a plasmonic lens. Nano Lett 5 (9):1726-1729. doi:Doi 10.1021/Nl051013j
23. López-Tejeira F, Rodrigo SG, Martín-Moreno L, García-Vidal FJ, Devaux E, Ebbesen TW, Krenn JR, Radko IP, Bozhevolnyi SI, González MU, Weeber JC, Dereux A (2007) Efficient unidirectional nanoslit couplers for surface plasmons. Nature Phys 3 (5):324-328. doi:10.1038/nphys584
24. Khoo EH, Guo Z, Ahmed I, Li EP (2012) Near-Field Switching and Focusing using Plasmonic Nanostructures with Different Polarizations. Proc SPIE 8457:1-8. doi:Artn 84573u

Doi 10.1117/12.929322

25. Steele JM, Liu ZW, Wang Y, Zhang X (2006) Resonant and non-resonant generation and focusing of surface plasmons with circular gratings. Opt Express 14 (12):5664-5670. doi:Doi 10.1364/Oe.14.005664
26. Fu YQ, Liu Y, Zhou XL, Xu ZW, Fang FZ (2010) Experimental investigation of superfocusing of plasmonic lens with chirped circular nanoslits. Opt Express 18 (4):3438-3443
27. Chen WB, Abeysinghe DC, Nelson RL, Zhan QW (2009) Plasmonic Lens Made of Multiple Concentric Metallic Rings under Radially Polarized Illumination. Nano Lett 9 (12):4320-4325. doi:Doi 10.1021/Nl903145p
28. Fang ZY, Peng QA, Song WT, Hao FH, Wang J, Nordlander P, Zhu X (2011) Plasmonic Focusing in Symmetry Broken Nanocorrals. Nano Lett 11 (2):893-897. doi:Doi 10.1021/Nl104333n
29. Gjonaj B, David A, Blau Y, Spektor G, Orenstein M, Dolev S, Bartal G (2014) Sub-100 nm Focusing of Short Wavelength Plasmons in Homogeneous 2D Space. Nano Lett 14 (10):5598-5602. doi:Doi 10.1021/Nl502080n
30. Martin PJ, Sainty WG, Netterfield RP (1984) Enhanced Gold Film Bonding by Ion-Assisted Deposition. Appl Optics 23 (16):2668-2669
31. Lerman GM, Yanai A, Levy U (2009) Demonstration of Nanofocusing by the use of Plasmonic Lens Illuminated with Radially Polarized Light. Nano Lett 9 (5):2139-2143. doi:Doi 10.1021/Nl900694r
32. Koev ST, Agrawal A, Lezec HJ, Aksyuk VA (2012) An Efficient Large-Area Grating Coupler for Surface Plasmon Polaritons. Plasmonics 7 (2):269-277. doi:DOI 10.1007/s11468-011-9303-7

33. Raether H (1988) Surface-Plasmons on Smooth and Rough Surfaces and on Gratings. Springer Tr Mod Phys 111:1-133
34. Johnson PB, Christy RW (1972) Optical Constants of the Noble Metals. Physical Review B 6 (12):4370-4379

**Figure Captions**

**Fig. 1** SEM micrograph of fabricated curved gratings for four different sector angles

**Fig. 2** (Schematic of NSOM experimental setup

**Fig. 3** NSOM scan (a) compared with simulation intensity map, (b) showing the focusing of coupled surface plasmons for $120°$ sector angle curved gratings

**Fig. 4** Intensity profile along the bisector (x axis in fig.6) of the simulated and experimental gratings (both with grating period of $\Lambda_g = 770$ nm) determined at a 10 nm vertical height from the gold-air interface

**Fig. 5** Measurement and simulation for the vertical width of the focal spot from simulation (blue diamonds) and experimental measurement (red circles), and reciprocal of the effective NA (black line) which is consistent with the experimental and simulation results

**Fig. 6** Schematic of curved grating with sector angle of $120°$. The linearly polarised light makes an angle of $\alpha$ at each point with the grating grooves

**Fig. 7** (a) Trench grating illuminated through the gold substrate (material with higher refractive index), (b) average intensity of coupled incident light into surface plasmons versus the depth of grooves (*DG*). Each colour symbol shows a different groove width (*GW*) as a ration of the grating period ($\Lambda_{gr}/m$)

**Fig. 8** Wavelength response of the curved grating from simulations – 30 nm thick gold film on glass with 120 degree sector angle

**Fig. 9** (a) 3D simulation of near field intensity map of $120°$ sector angle curved grating with optimized structural parameters (*GW*= 385 nm, and *GD*=120 nm) illuminated at $\lambda_0 = 790$ nm, (b) The intensity profile along the bisector of the grating ( x axis in fig. 6)

**Fig. 10** Schematic for two faced curved gratings centred at the same point and in a symmetry-broken configuration

**Fig. 11** Simulation of the performance of single curved grating and asymmetric double sided curved gratings with six grooves

**Fig. 12** Comparison of the longitudinal intensity profile along the bisector of the single curved grating and asymmetric double-sided curved gratings with six grooves

**Figures;**

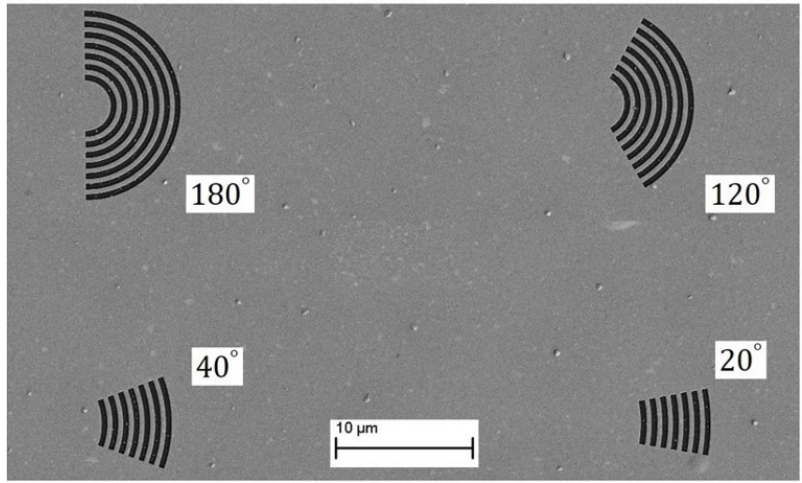

**Fig. 1**

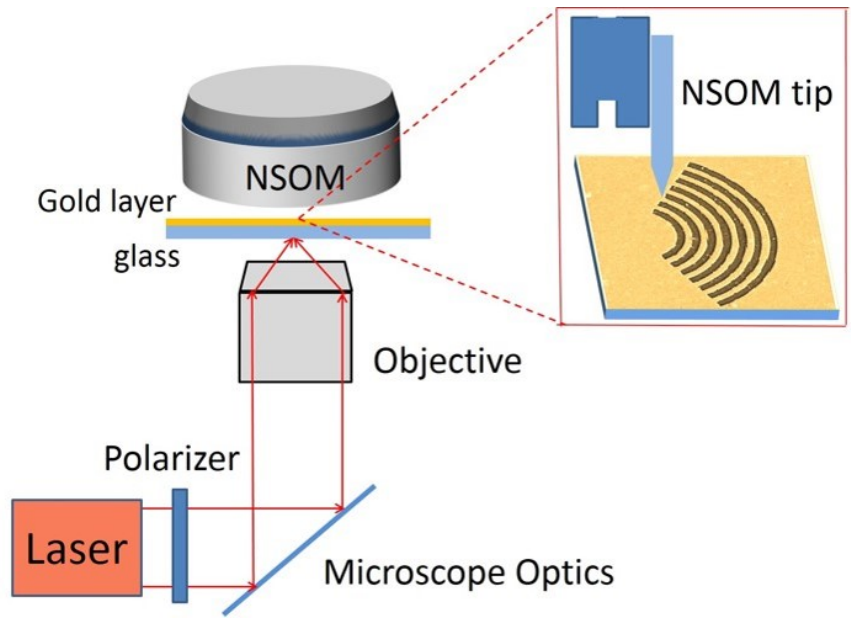

**Fig. 2**

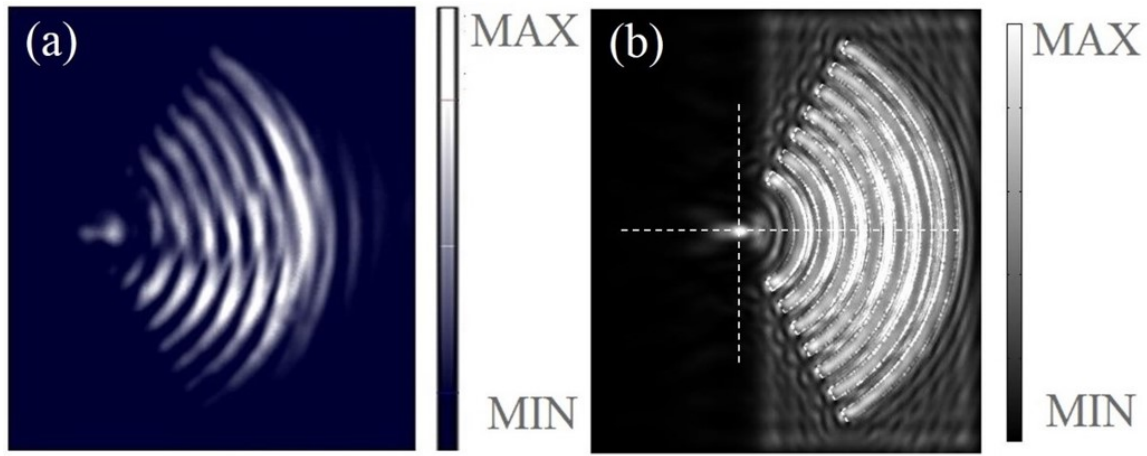

**Fig. 3**

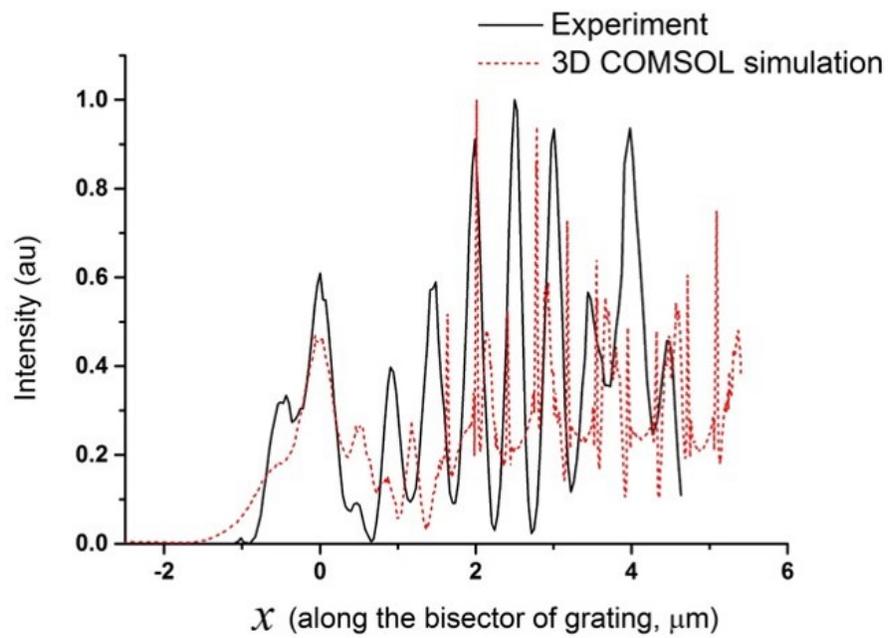

**Fig. 4**

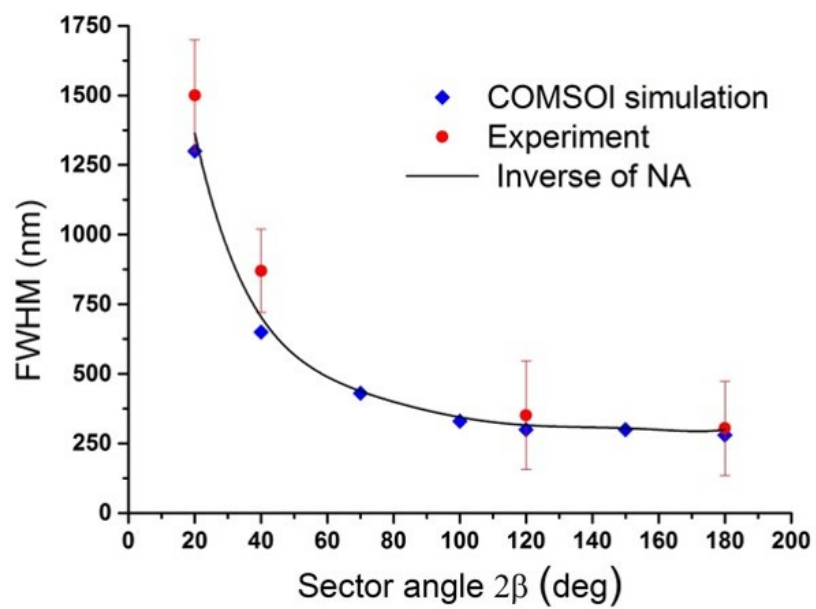

**Fig. 5**

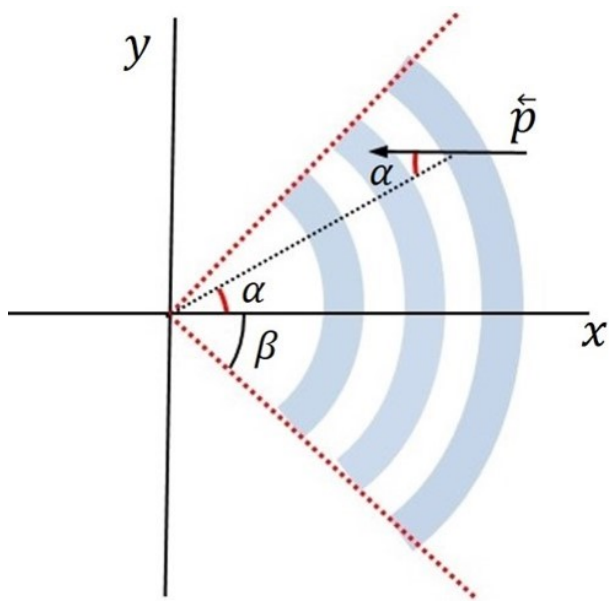

**Fig. 6**

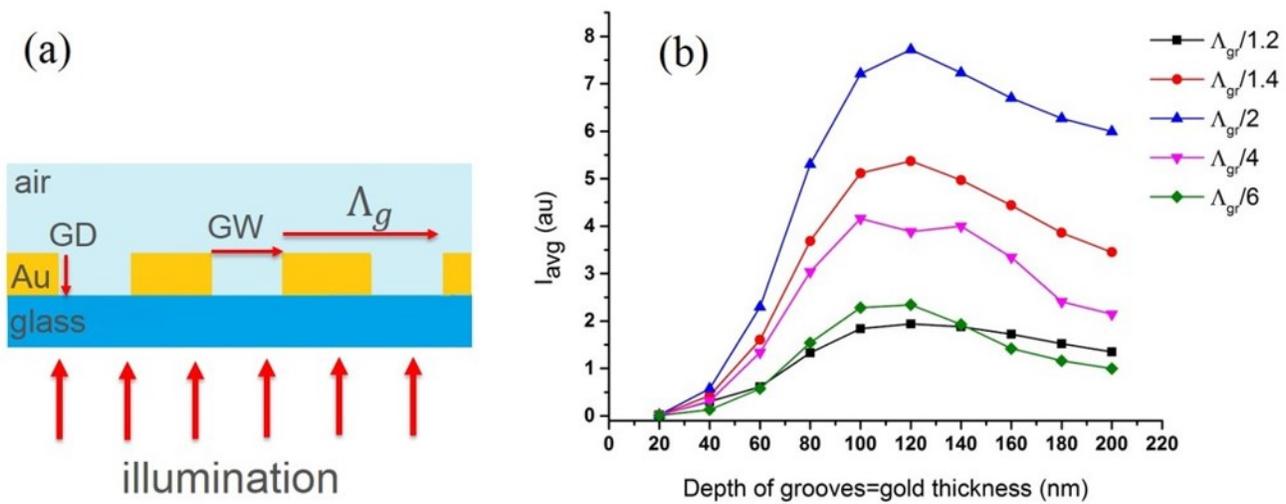

**Fig. 7**

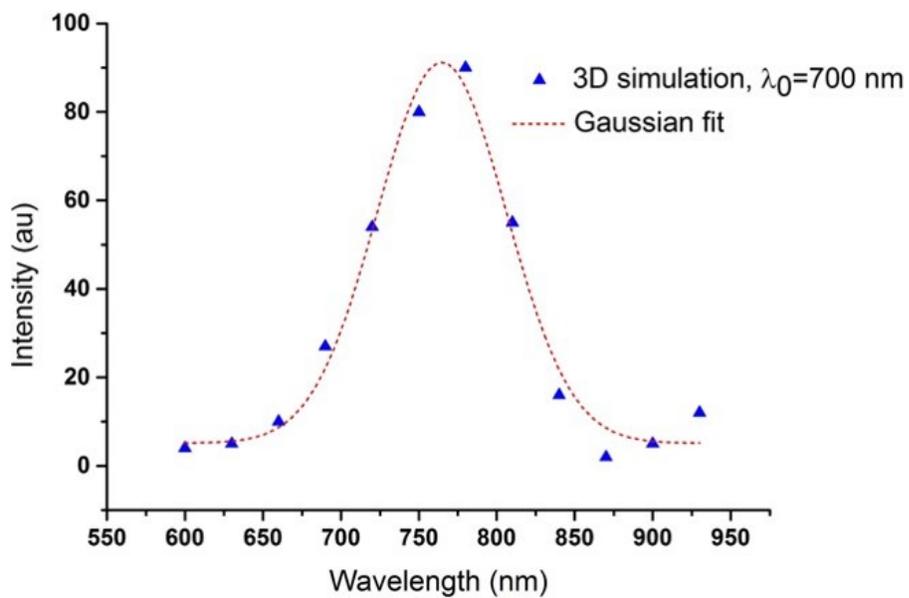

**Fig. 8**

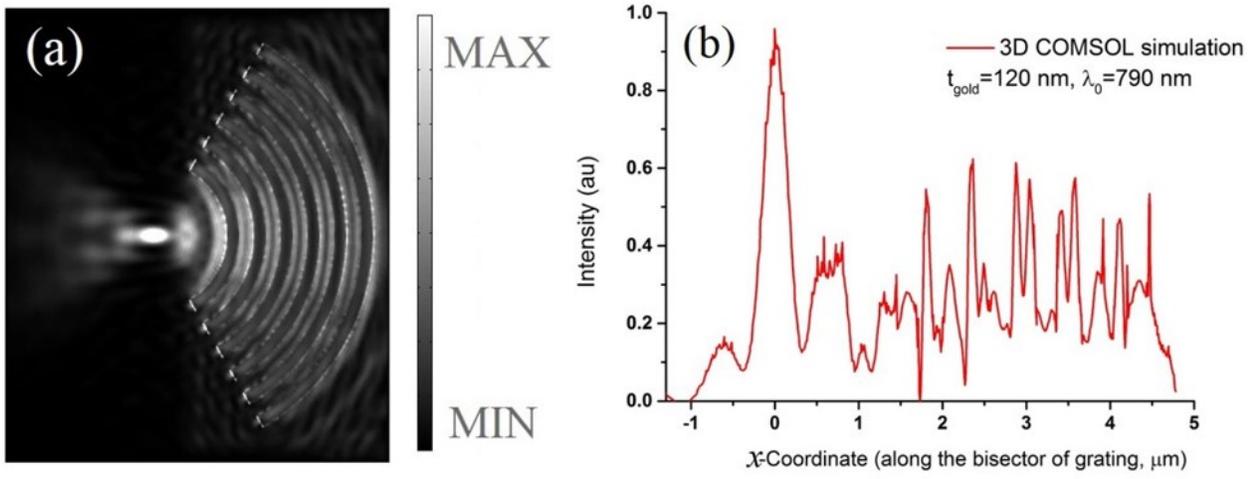

**Fig. 9**

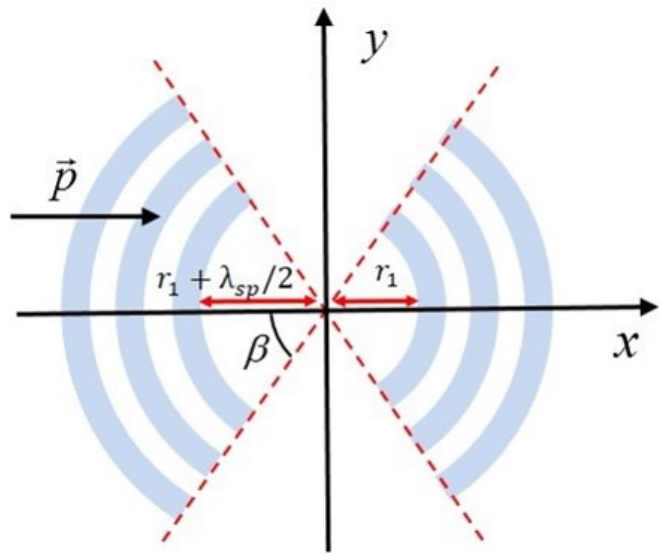

**Fig. 10**

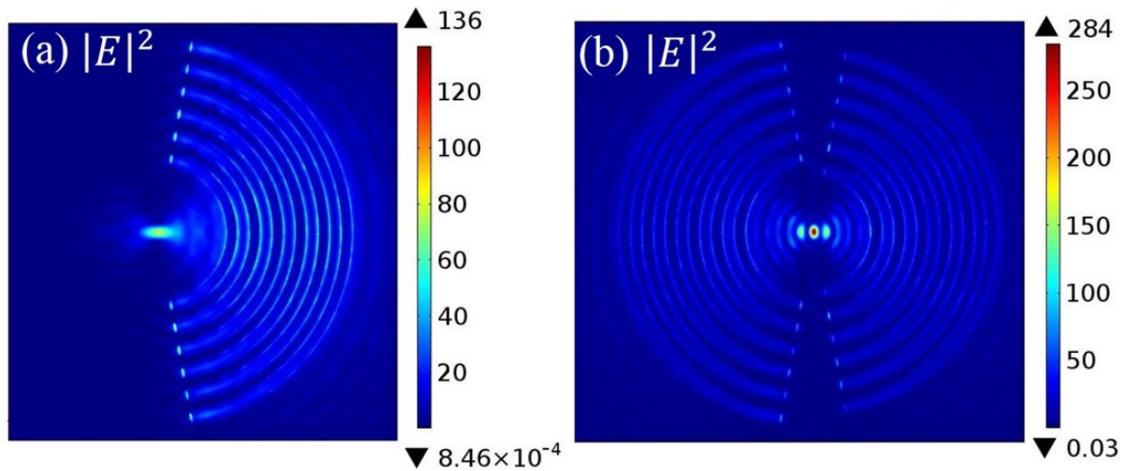

**Fig. 11**

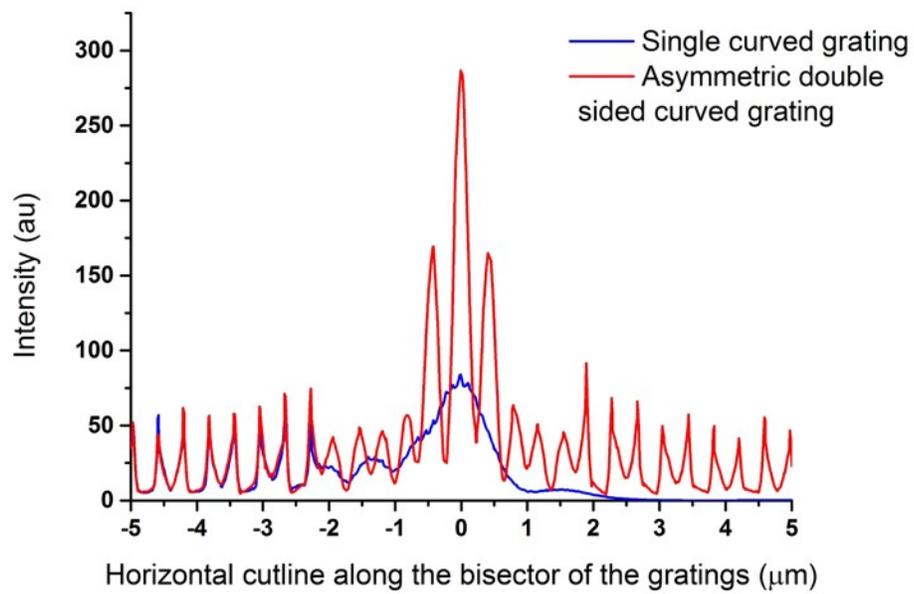

**Fig. 12**